# Modification of Excitonic Emission in GaN Microcavities Formed by Planar Hexagonal Micro-Nuts Grown by Selective Area Epitaxy

Galia Pozina[1*], Alexey V. Belonovski[2], Iaroslav V. Levitskii[3,4], Maxim I. Mitrofanov[3,4], Elizaveta I. Girshova[2,3,4] Konstantin A. Ivanov[3], Gleb V. Voznyuk[3], Sergei N. Rodin[4], Konstantin M. Morozov[2], Vadim P. Evtikhiev[4], Mikhail A. Kaliteevski[2,3,4]

[1] Department of Physics, Chemistry and Biology (IFM), Linköping University, S-581 83 Linköping, Sweden
[2] St-Petersburg Academic University, Khlopina 8/3, 194021 St. Petersburg, Russian Federation
[3] ITMO University, Kronverkskiy pr. 49, 197101 St. Petersburg, Russian Federation
[4] Ioffe Institute, Politekhnicheskaya 26, 194021 St. Petersburg, Russian Federation

* galia.pozina@liu.se

## Abstract

GaN-based resonators supporting whispering gallery modes are suggested as a promising approach for fabrication of efficient nanophotonic applications. A modification of excitonic emission due to coupling with cavity modes is observed experimentally in the GaN planar hexagonal microcavities grown by a selective area metal-organic vapor phase epitaxy. Low temperature cathodoluminescence spectra measured in-situ scanning electron microscope show two or three emission peaks split by 100 meV or 160 meV for the resonators formed by double- and single-wall micro-nuts, respectively. The results are compared with numerical calculations of the cavity mode energies, their Q-factors and the spatial distribution of the modes' intensities. Although most modes are characterized by a small Purcell coefficient, there are a limited number of isolated cavity modes with a large Purcell factor. Such modes can efficiently interact with the bulk exciton in GaN. The results are also analyzed by diagonalization of Hamiltanion describing the system of interacting exciton and cavity modes. Theoretically, it is shown that a strong coupling between the exciton and the cavity modes is possible in such micro-resonators.

## Introduction

III-nitride semiconductor nanostructures attract significant interest due to diverse applications in many fields of science and technology covering optoelectronics, nanophotonics, biology and medicine [1-3]. Technology applied for fabrication of optoelectronic and electronic devises based on nanostructures requires a firm capability to control morphological, structural and electronic properties of III-N nanostructured material. The manufacturing process involves usually epitaxial growth of multilayers on substrate [4] with postgrowth lithography and processing, so-called top-down paradigm [5]. An alternative bottom-up method [6] for fabrication of III-N nanostructures includes a selective area epitaxy [7,8] or the growth techniques based on processes with self-organized formation of nanostructures [9,10]. A promising method to produce nanostructures with controllable sizes and shapes is a selective area metal-organic vapor phase epitaxy (MOVPE), where a given pattern in a mask can be etched by focused ion beam (FIB) [11,12]. One of the important nanophotonic challenges is related to design of microcavities for efficient emission. There is a large number of possible resonator schemes including microcavities in the form of micropillars, micro-discs [13,14] or hexagonal nano rods [15]. Latter can be fabricated without the Bragg reflectors while supporting whispering gallery modes (WGMs) and, thus, demonstrating an enhancement of the spontaneous emission rate, i.e. so-called the Purcell effect [16]. The approach is very attractive since a coherent growth of multilayered Bragg reflectors based on III-nitrides is difficult and yet need to be developed since the growth rate of III-nitrides is strongly dependent on temperature of synthesis, additionally, the III-nitride-based quantum wells exhibit a rather broad linewidths and are influenced by the quantum-confined Stark effect if grown in polar [0001] direction [17]. On the other hands, III-N-based microcavities supporting WGM can demonstrate a significant enhancement of the spontaneous emission rate even for rather low Q-factors [18,19]. Attempts to use wide band gap III-nitride semiconductors for fabrication of microcavities are justified in terms of material properties such as large oscillator strength and large exciton-binding energy (~25 meV in bulk GaN), which paves the way for efficient performance of optoelectronic devices even at high temperatures [20, 21].

An important phenomenon observed in semiconductor microcavity is strong coupling of exciton and photon modes [22]. The bulk GaN has 3 types of exciton (usually referred as A, B, and C-exciton), and A- exciton is GaN possesses large longitudinal-transverse splitting $\omega_{LT}$ of the order of 1 meV [23]. Since the value of Rabi splitting $\Delta\omega$ of exciton and photon modes in microcavity is about $\Delta\omega \sim \sqrt{\omega_{LT}\,\omega_0}$, where $\omega_0$ is the frequency of the cavity mode, the value of Rabi splitting in III-N based microcavities exceeds 50 meV [24] and make it possible to observe the polariton lasing at room temperature [25].
Thus, in III-N based microcavities, the light-matter interaction can lead to a strong coupling regime with appearance of pronounced exciton-polariton modes.

The purpose of this work is to report on a recent progress in development of GaN microcavities designed in the form of planar hexagonal micro-nuts. Spatially-resolved cathodoluminescence (CL) measurements reveal in such micro-resonators a modification of excitonic emission explained theoretically in terms of a strong coupling regime between WGMs and excitons resulting in a huge Rabi splitting exceeding 100 meV.

**Results**

The GaN micro-resonators have been fabricated by a selective area MOVPE employing FIB etching for pattering the mask layer. Details of the growth, process parameters and characterization equipment are described in Materials and Methods. The pattern in the form of equidistant rings with a diameter of 5 µm was applied (the rim width was ~100 nm). The etched ring after this process step is illustrated in Figure 1a. Formation of GaN hexagons in the etched pattern is illustrated in Figure 1b and 1c. The hexagon wall growth starts with a nucleation of small pyramids in the etched ring (Figure 1b), then, these pyramids enlarge and form a faceted structure (Figure 1c). The emerging faces of the microstructure are formed by semi-polar crystallographic planes such as $(10\bar{1}2)$, $(10\bar{1}1)$, $(20\bar{2}1)$, $(11\bar{2}2)$, as shown in Figure 1d. Finally, merging pyramids form a final hexagonal structure schematically shown in Figure 1e.

The optimization of selective area MOVPE process steps resulted in a coherent growth of planar GaN microcavities with two specific shapes as shown in scanning electron microscopy (SEM) images in Figure 2a, 2c and 2b, 2d, respectively. The difference in the shapes, i.e. between the single and the double hexagonal micro-nuts has been obtained by a slight difference in the etching pattern, namely, in the first case, the etching depth of a single ring corresponded to the thickness of the $Si_3N_4$ mask layer, while for the second type of hexagons, the etching of the same rings was done across the mask into the GaN buffer layer, as illustrated in Figure 3.

The ion beam used for etching has Gaussian-like spatial profile. Uniform shallow etching can be achieved by scanning of the low current ion beam with a small diameter along the mask layer, as illustrated in Figure 3a. In contrast, if the ion beam current is high then the beam diameter is large resulting in etched stripes of V-shape, as illustrated in Figure 3d. The two shapes of the stripe define two different growth modes. In the shallow flat stripe, the initial single pyramid is formed with the base parallel to the plane (0001) and facets oriented along semi-polar crystallographic planes such as $(10\bar{1}2)$, $(10\bar{1}1)$, $(20\bar{2}1)$, $(11\bar{2}2)$. When the base of the pyramid reaches the edge of the mask (Figure 3b), it starts to grow over the mask and forms the flat top due different growth rate in different crystallographic directions. In the V-shaped stripe, there are formations of multiple nucleus on the inclined planes (Figure 3e), they expand and reach edges of the mask faster than in the case of the flat shallow stripe. Then, the growth areas near the edges of the etched stripe consume the flow of precursors, which leads to the conservation of the gap between inner and outer hexagon (Figure 3f) and to the formation of the peculiar double hexagon shape. Note, that the groove between inner and outer hexagons has a faceted structure (Figure 2d).

In both cases, GaN hexagons are equidistant, have nearly perfect geometrical shapes and the same lateral size of ~7 μm. The wall thickness is ~1 μm and ~500 nm for the single and for the double hexagonal structures, respectively. The height of both microcavity types is ~1 μm as shown in the insets of Figure 2c and 2d. In such structures the round trip of light corresponds to approximately $10^2$ wavelengths of light at exciton frequency.

The panchromatic CL images (PCL) in Figure 4 give an overview of the emission intensity distribution within the GaN hexagonal microcavities. Figure 4a and 4c show CL data obtained at room temperature, while Figure 4b and 4d show CL images measured at 10 K. Clearly, a bright contrast in the CL maps corresponding to the higher intensity of emission is concentrated along the edges of the microcavity walls. In the double-wall hexagonal micro-nuts (Figure 4c and 4d), the CL images show a more complicated pattern with the bright contrast observed at the edges of both the inner and outer walls. Such spatial localization of the emission observed in the single- and double-wall micro-nuts even at room temperature is a strong evidence of the whispering gallery modes in the GaN hexagonal microcavities.

Further, to study optical properties of the hexagonal microcavities, we have investigated CL spectra at room temperature and at 10 K measured with a spatial resolution of ~100 nm, i.e. when the electron beam has been focused to a specific spot on the GaN microcavity. CL spectra were taken at different points as indicated in the insets of Figure 5a, 5b, 5c, and 5d, respectively. For reference, CL spectra measured at similar conditions at room temperature and at 10 K for the GaN epitaxial layer are shown by the dashed lines in Figure 5a, 5b and 5c, 5d, respectively. We note that all GaN hexagons of each type have demonstrated at 295 K a rather broad emission peaking at ~3.3 eV, though some small variations of the energy position have been observed. The emission band maxima of the GaN hexagons are noticeably shifted to lower energies compared to the GaN layer. In latter case, the CL line is peaking at ~3.4 eV at 295 K, which corresponds to a typical excitonic emission in GaN [26]. The full width at the half maximum (FWHM) for the CL bands measured for the GaN hexagonal microcavities is in the range of 200-300 meV at room temperature, which is much broader than the FWHM of ~60 meV for the excitonic CL spectrum taken at the GaN layer. There is a feature at ~3.4 eV at the higher energy shoulder of the CL bands measured at point 1 and 2 for both types of microcavities (Figure 5a and 5b) indicating that there is an overlapping between the GaN exciton line and some addition emission, whose origin can be related to the microcavity modes.

While the room temperature CL spectra are very similar for all measured points, the low temperature results have revealed a large variety in the emission spectra shape depending on the measured point at the hexagon GaN microcavities. Some examples are collected in Figure 5c and 5d for the single- and double-wall microcavities, respectively. For comparison, the near-band-gap emission with the peak at ~3.46 eV taken at 10 K for the GaN layer is shown by the dashed lines. The 3.46 eV CL line is related to the exciton bound to shallow donors (DBE) such as silicon and oxygen [27]. Obviously, the CL spectra for GaN hexagonal microcavities are very different having a complex shape consisting of two and three bands for the single and double-wall microcavities, respectively.

The relative intensity between the bands vary essentially depending on the acquisition point. The bands, however, have a rather firm energy separation for each microcavity type: ~160 meV and ~100 meV for the single and double hexagons, respectively (see dash-dotted lines in Figure 5c and 5d). The energy positions for the emission maxima are following: ~3.30 and 3.46 eV for the single and 3.23, 3.33 and 3.43 eV for the double hexagonal structures, respectively. The FWHM of the emission lines at 10 K is in the range of 80-90 meV as estimated from the Gaussian fitting. These widths exceed the FWHM of ~50 meV for the DBE line in the GaN layer.

The standard way of the analysis of the polariton behavior in microcavities is studying of the polariton dispersion via angle-resolved measurements. In our case, where cavity modes experience confinement in all 3 dimensions, such method cannot be utilized. At the same time, detuning between the exciton and the photon modes can be changed using different approach: by changing the temperature of the system. The energies of excitons vary with the change of temperature in line with the value of the band gap energy, while the variation of frequency for the cavity modes is governed by temperature dependence of refractive index. Figure 6a shows photoluminescence (PL) spectra of the single hexagon structure taken at different temperatures varying from 5 K to room temperature. The increase of the temperature leads to the red shift of the higher energy peak. The lower energy peak also experiences the reduction of the emission energy, but as it is clear from Figure 6b, the red shift is much smaller for the lower energy peak. Such behavior indicates that the lower energy emission peak is associated with the cavity mode.

**Discussion**

Thus, the following main results have been observed for the planar GaN hexagons:

i) the spectral position of the emission peaks does not depend on the point of excitation of the sample.

ii) the energy interval between the emission peaks is of the order of 100 meV

iii) at low temperature, there are two (three) peaks in the luminescence spectra for the single (double) hexagonal structures, however, when temperature raises, the separated peaks merge in the one wide emission band.

Such behavior of the emission in GaN microcavities indicates the strong modification of the excitonic emission from the structures by cavity modes, which can occur in weak or strong coupling regime. The shape of the dependence of the peak's energies shown in Figure 6b can be considered as a part of anti-crossing between exciton and cavity modes (though with the range of detuning occurring with temperature variation from 5 K to room temperature anti-crossing region is not achieved). With increasing temperature, the two peaks merge (note, that in Figure 5 spectra are shown in linear scale while in Figure 6 in semilogarithmic scale).

The luminescence spectra demonstrate polaritonic nature of the emission in the studied GaN hexagonal structures. Consequently, we analyze theoretically exciton-cavity interaction to reveal if such interaction can occur in such structures in the strong coupling regime despite a large size of systems and a high density of cavity modes. For this purpose, we model the mode structures of hexagonal cavities and their interaction with a bulk exciton in GaN.

We note that a rigorous three-dimensional (3D) modeling of the optical mode structures of the studied hexagonal microcavities is a cumbersome task (since the size of the structure is about 5 μm x 5 μm, and the wavelength of light is about 150 nm), which is beyond the scope of this paper. At the same time, important conclusions related to optical modes can be based on modeling of two-dimensional (2D) hexagons as shown in Figure 7. Cavity mode energies, their Q-factors, and spatial distribution of the cavity modes intensity was made by solving the Maxwell's equations using numerical calculations [28] for ideal single and double hexagons, respectively. For modeling in the case of single hexagon resonator, the size was chosen as 1.95 μm and 3.20 μm for the inner and outer sizes (i.e. radii), respectively. For the double-wall micro-nut resonator, the following parameters have been used: the inner hexagon radius - 3.46 μm, the circle radius - 2.67 μm, the intermediate hexagon radius - 2.4 μm and the radius of the inner hexagon -1.62 µm. The details of numerical modeling and the model resonators can be found in Materials and Methods.

There are many different modes in the considered microcavities with an interval of 28 meV. The modes have different spatial distribution of the intensity of the light wave and different Q-factors. Figures 7a and 7b show examples of electromagnetic field distribution calculated for the single hexagon microcavity for the modes with the energy 3.26 eV and 3.3 eV, respectively. These energies are close to the experimental CL band energy (~3.3 eV). Figures 7c and 7d show examples of the mode intensity distribution in the outer hexagon for the mode energy of 3.35 eV and for the inner hexagon for the mode energy of 3.37 eV. It is clear that different modes can be strongly (Figures 7a and 7d) or weakly (Figures 7b and 7c) localized.

Purcell coefficient $F_p$, describing the strength of interaction of the cavity modes with an emitter reads as:

$$F_p \approx \frac{Q\lambda^3}{n^3 V} \quad (1)$$

where $\lambda$ is the wavelength of light corresponding to the frequency of the mode, $n$ is refractive index and $V$ is the volume of the mode [14,29,30].

For simplified analysis of 2D case, we calculate the quantity $F$, defined as

$$F = \frac{Q\lambda^2}{n^2 S} \quad (2)$$

where $S$ is the effective area of the mode in the cavity.

Figures 8a and 8b show the energies and quantities $F = \frac{Q\lambda^2}{n^2 S}$ for the eigen modes localized in the single and double hexagonal structures, respectively. Clearly, while most of the modes are characterized by moderate values of $F$, there are still few modes with a rather high $F$. Figures 8c and 8d show the distribution of the mode probability depending on $F$ for single and double hexagonal structures, respectively. For the single hexagonal structure, almost all modes are characterized by the value of $F$ not exceeding 50. For the double hexagonal structure, the value of $F$ does not exceed 5 for most of the cavity modes. At the same time, there is a small probability of appearance of modes with a high value of $F$ for both types of the hexagonal structures. For the 3D case, it is expected that a qualitative behavior of the modes will be similar, namely, most of the modes will be characterized by a small Purcell coefficient $F_p$, while a limited number of isolated cavity modes will have a high $F_p$ and will efficiently interact with a bulk exciton in GaN. For optical modes with high Q-factors localized within hexagons, only a small exponentially decaying tail of the electromagnetic field penetrates into the doped buffer layer. Thus, doping-related absorption of light only slightly reduces Q-factor of such modes. In contrary, for the modes with low Q-factors penetrating into the substrate, the absorption in the doped buffer layer reduces Q-factor further, which decreases the strength of interaction between the exciton and the optical mode, thus, contributing to the selection of the modes interacting with the exciton.

The result of such interaction can be described as follows. The Hamiltonian of the interaction for the system consists of bulk exciton $|x\rangle$ and N modes of the cavity $[C_1, ..., C_2]$ can be simplified by using rotating wave approximation. In the case of exciton Hamiltonian, we will exploit second quantized form and field operators. The total Hamiltonian reads as:

$$\mathcal{H} = \hbar\omega_0 \hat{x}^+ \hat{x} + \sum_k \hbar\omega_k \hat{c}_k^+ \hat{c}_k + \sum_k \hbar(g_k \hat{c}_k \hat{x}^+ + g_k^* \hat{c}_k^+ \hat{x}), \qquad (3)$$

where $\omega_0$ is the exciton frequency, $\omega_k$ are the frequencies of cavity modes, $g_k$ are the constants describing exciton-photon interactions, and $\hat{x}, \hat{c}_k$ ($\hat{x}^+, \hat{c}_k^+$) are the operators of annihilation (creation) of exciton and photons, respectively.

The Hamiltonian (3) can be rewritten in the matrix form

$$\mathcal{H} = \hbar[\hat{x}^+ \; \hat{c}_1^+ \; \hat{c}_2^+ \cdots \hat{c}_N^+] \cdot \begin{bmatrix} \omega_0 & g_1 & g_2 & \cdots & g_N \\ g_1^* & \omega_1 & 0 & \cdots & 0 \\ g_2^* & 0 & \omega_2 & \cdots & 0 \\ \vdots & \vdots & \vdots & \ddots & \vdots \\ g_N^* & 0 & 0 & \cdots & \omega_N \end{bmatrix} \cdot \begin{bmatrix} \hat{x} \\ \hat{c}_1 \\ \hat{c}_2 \\ \vdots \\ \hat{c}_N \end{bmatrix}. \quad (4)$$

The procedure of diagonalization the Hamiltonian (4) is described in Materials and Methods. Eigenvalues $\lambda_i$ of the matrix in (4) correspond to the energies of the N+1 exciton-polariton modes $|i\rangle$, which are the eigenstates of the Hamiltonian (4). Thus, we can write

$$\mathcal{H}|i\rangle = \sum_k \hbar \lambda_k \hat{p}_k^+ \hat{p}_k |i\rangle = \hbar \lambda_i |i\rangle, \quad (5)$$

where

$$\hat{p}_k^+ = a_{0k}\hat{x}^+ + a_{1k}\hat{c}_1^+ + a_{2k}\hat{c}_2^+ \ldots a_{Nk}\hat{c}_N^+ \quad (6)$$

and $a_{0k}$ ($a_{1k}, a_{2k}, \ldots a_{Nk}$) are the weight coefficients of exciton (cavity modes) in polariton modes.

The strength of interaction of exciton with specific mode $g_k$ is proportional to quantity $F_k$:

$$g_k \sim F_k W_0, \quad (7)$$

where $W_0$ is the probability of excitonic emission in the uniform media and coefficient $F_k$ is defined by Equation 2.

As seen in Figures 8a, 8b, 8c, 8d, most of the modes are characterized by small $F$ of the order of unity, while there could be the cavity modes interacting to the exciton with both large value of $F$ and strength of interaction $g_k$. The intensity of the excitonic emission associated with specific polariton mode $|i\rangle$ will correlate with its excitonic fraction described by $|a_{0i}|^2$.

Using a simplified model that could explain the observed results of cathodoluminescence for GaN hexagonal microcavities, we analyzed the structure of the polaritonic mode under the assumption that the interaction of the exciton with all but one cavity modes is weak and is characterized by the interaction strength g = 4 meV. However, for one particular cavity mode with an energy of 3.307 eV, the interaction strength was chosen to be g = 30 meV.

Figure 8e shows the dependence of the excitonic contribution to the polaritonic state $|a_{0i}|^2$ as a function of the energy of the polariton mode for the single hexagon microcavity. Clearly, two peak behavior is mimicking the experimental two peaks observed in low temperature luminescence spectra, thus, illustrating a strong coupling between the exciton and the cavity mode.

Similarly, Figure 8f shows the dependence of the excitonic contribution $|a_{0i}|^2$ calculated for the double hexagonal structure. Parameters used for modelling were similar to the previous, namely, for all cavity modes (except two), the strength of interaction was chosen to be weak g = 4 meV, while for the mode with an energy of 3.24 eV we used the strength of interaction g = 40 meV and for the cavity mode with an energy of 3.34 eV the strength of interaction was g = 30 meV. One can see that a three peak behavior is similar to the experimentally observed luminescence spectra at low temperature for the double-wall hexagonal microcavity.

In summary, GaN microcavities have been fabricated by a selective area MOVPE using FIB for etching of the pattern in the $Si_3N_4$ mask. We have studied two types of the microcavities: the single wall and the double walls GaN hexagons. The shape of the microcavities has been checked by SEM and it was found that the hexagons were coherently grown and almost identical. The optical properties of the microcavities have been controlled using low temperature CL measurements with a nanoscale spatial resolution. It was obtained that at low temperature the CL spectra can consist of two and three near-band-gap emission lines for the single and double resonator, respectively. The emission bands have different relative intensities depending on the measured points. At room temperatures, due to broadening, different bands were not resolved and, thus, merge to a broad single emission. Numerical simulations of the cavity mode for the hexagon resonators with parameters similar to experimental have shown that some particular cavity modes can be more localized compared to others. Such localized cavity modes could interact with an exciton in a weak or strong coupling regime. The analysis of the exciton-polariton modes carried out by means of quantum electrodynamics provides possible explanation for the observed emission spectra in GaN planar microcavities.

**Materials and methods**

*Growth of GaN planar hexagonal microstructures*

The GaN microresonators were fabricated as following. First, a 3-µm-thick GaN buffer layer was grown on a (0001) sapphire substrate by metal-organic vapor phase deposition (MOVPE). The buffer layer was doped by silicon with a concentration of $2 \times 10^{17}$ cm$^{-3}$. The doping of the buffer layer is necessary to prevent charging, which can deflect and defocus the ion beam during a focused ion beam (FIB) patterning of the mask. A silicon nitride ($Si_3N_4$) mask was deposited on the top of the buffer layer. Trimethylgallium (TMG), silane ($SiH_4$) and ammonia ($NH_3$) were used as precursors. The mask layer was deposited at the growth temperature of 1000 °C. Then, a pattern in the mask has been etched. For the FIB etching, we used a Ga ion beam with energy of 30 keV and with a probe current of 450 pA. The pattern has the form of equidistant rings with the diameter of 5 µm and the rim width of ~100 nm. Depending on the depth of etching we can get single-wall or double-wall hexagons. For the first type of microstructures we etched only the mask layer while for the second type of hexagons, the etching was deeper, i.e. inside the buffer GaN layer. To prevent the formation of Ga droplets on the mask surface, a xenon difluoride ($XeF_2$) was directed to the etching area. The windows in the mask can be etched with a spatial resolution of ~ 5 nm. After the FIB processing, the MOVPE growth was carried out to fabricate GaN microcavities. Precursors' flows of 200 cm$^3$/min and 60 µmol/min have been used for ammonia and TMG, respectively. Hydrogen was used as a carrier gas and its flow was 6000

cm$^3$/min. The pressure during the growth was 100 mbar and the temperature of the substrate was kept to 1030 °C.

The grown structures were studied using a LEO 1550 Gemini scanning electron microscope (SEM) combined with MonoCL4 system with a liquid-He-cooled stage for low-temperature cathodoluminescence (CL) measurements. Acceleration voltage of the electron beam was 5 kV, which allows acquisition of CL images and spectra with nanoscale spatial resolution. A fast CCD detection system and a Peltier cooled photomultiplier tube have been used for acquisition of CL spectra and for CL mapping, respectively. PL spectra have been measured using micro-photoluminescence set-up with a spatial resolution of ~1 µm. The third harmonics ($\lambda_e$ = 266 nm) from a Ti:sapphire femtosecond pulsed laser with a frequency of 75 MHz has been used as an excitation source. The samples were placed inside a variable temperature (5-300 K) Oxford Microstat allowing X−Y translation with a high precision better than 0.5 µm.

*Numerical modeling of cavity modes*

Microcavities eigenmodes study was carried out in two dimensions in Electromagnetic Wave, frequency domain module of COMSOL Multiphysics software. The full size of the model is 10.2x10.2 µm for both single and double wall hexagonal microcavities simulations. Based on SEM pictures of the microcavities we approximated the geometry for the simulations as follows: for single wall hexagonal microcavity – 3.20 and 1.95 µm for outer and inner radii respectively, for double wall hexagonal microcavity – 3.46, 2.67, 2.40 and 1.62 µm for outer and inner radii of the respective hexagon. The inner boundary of the larger/outer hexagon was approximated as a circle since it was not exactly pronounced from the SEM picture. The microcavities geometry is depicted in Figure 9. In our simulations we considered that microcavities are made of material with no absorption and refractive index $n = 2.6267$ [31].

The ambient media has refractive index $n_0 = 1$. The boundaries of the model consist of perfectly matched layers (PML) to simulate open boundaries. Perfect magnetic conductor (PMC) boundary condition was applied on the perimeter of the model. In the model we used free triangular mesh with maximum element size of 27 nm for microcavity and 74 nm for ambient media. MUMPS (MUltifrontal Massively Parallel sparse direct Solver) was used as an eigenvalue solver.

*Diagonalization the Hamiltonian of the interaction*

The Hamiltonian of the interaction in the matrix form reads as:

$$\mathcal{H} = \hbar[\hat{x}^+ \ \hat{c}_1^+ \ \hat{c}_2^+ \cdots \hat{c}_N^+] \cdot \begin{bmatrix} w_0 & g_1 & g_2 & \cdots & g_N \\ g_1^* & w_1 & 0 & \cdots & 0 \\ g_2^* & 0 & w_2 & \cdots & 0 \\ \vdots & \vdots & \vdots & \ddots & \vdots \\ g_N^* & 0 & 0 & \cdots & w_N \end{bmatrix} \cdot \begin{bmatrix} \hat{x} \\ \hat{c}_1 \\ \hat{c}_2 \\ \vdots \\ \hat{c}_N \end{bmatrix} = \hbar \hat{C}^+ W \hat{C} \qquad (8)$$

where $C^+$ is a row, made up of operators of creation, $C$ is a column, made up of operators of annihilation, W is a square matrix.

The square matrix W is Hermitian, therefore, its diagonalization is possible with the help of a unitary transformation:

$$A^{-1} W A = \lambda \qquad (9)$$

where $A$ is unitary matrix whose columns contain the eigenvectors of the matrix $W$, $\lambda$ is diagonal matrix whose elements are the eigenvalues of the matrix $W$. Insertion of this result into Eq. (8) produces

$$\mathcal{H} = \hbar \cdot \hat{C}^+ A \cdot \lambda \cdot A^{-1} \hat{C}. \qquad (10)$$

Expression $\hat{C}^+ A$ gives a new form of operators of creation in a mixed system «exciton+modes». $A^{-1}\hat{C}$ is a new form of operators of annihilation. Consider this expression:

$$\hat{C}^+ A = [\hat{x}^+ \ \hat{c}_1^+ \ \hat{c}_2^+ \ \cdots \ \hat{c}_N^+] \begin{bmatrix} a_{00} & a_{01} & a_{02} & \cdots & a_{0N} \\ a_{10} & a_{11} & a_{12} & \cdots & a_{1N} \\ a_{20} & a_{21} & a_{22} & \cdots & a_{2N} \\ \vdots & \vdots & \vdots & \ddots & \vdots \\ a_{N0} & a_{N1} & a_{N2} & \cdots & a_{NN} \end{bmatrix} =$$

$$= \left[\hat{x}^+ a_{00} + \sum_{j=1}^{N} \hat{c}_j^+ a_{j0}, \hat{x}^+ a_{01} + \sum_{j=1}^{N} \hat{c}_j^+ a_{j1}, \cdots, \hat{x}^+ a_{0N} + \sum_{j=1}^{N} \hat{c}_j^+ a_{jN}\right] =$$

$$= [\hat{p}_0^+ \ \hat{p}_1^+ \ \cdots \ \hat{p}_N^+] = \hat{P}^+ \tag{11}$$

where $\hat{p}_k^+ = \hat{x}^+ a_{0k} + \sum_{j=1}^{N} \hat{c}_j^+ a_{jk}$ is a new form of operator of creation. $\hat{P}^+$ is a row, made up of new operators of creation. Similar transformations can be carried out with another expression $A^{-1}\hat{C} = \hat{P}$. Here $\hat{P}$ is a column made up of operators of annihilation.

In the new notation, the Hamiltonian (S3) will take the form:

$$\mathcal{H} = \hbar \ P \ \lambda \hat{P}^+ = \sum_k \hbar \lambda_k \hat{p}_k^+ \hat{p}_k \tag{12}$$

Thus, we obtained a diagonalized Hamiltonian with new operators $\hat{p}_k^+$ and $\hat{p}_k$. $\lambda_k$ is the frequency correspond to the energies of the N+1 exciton-polariton modes.

Data availability. All data generated and/or analysed during this study are available from the corresponding author on reasonable request.

**Acknowledgements**

The work has been supported by Russian Science Foundation Grant 16-12-10503 and Swedish Energy Agency.

**Conflict of interest**

The authors declare that they have no conflict of interest.

**Authors contributions**

GP, MAK and VPE designed the research idea. MIM, IVL, GVV, VPE and SNR contributed to FIB and MOVPE. GP performed optical and SEM experiments. AVB, IVL, MAK, EIG, KAI, KMM carried out numerical calculations and modeling. All authors contributed to discussion and analysis of the results. GP, MAK, AVB wrote the manuscript with contributions from all authors.

**Figure legends**

Figure 1. SEM images of the different process steps showing formation of GaN microcavity. (a) The pattern in the $Si_3N_4$ mask in the form of the single ring. (b) After a short MOVPE growth run: a nucleation of the small GaN pyramids can be detected within the etched ring. (c) After a longer MOVPE growth: GaN pyramids enlarged and start to form faceted walls. (d) Schematic drawing of the wurtzite unit cell, low index facet $(0\bar{1}11)$ is shown. (e) Schematic drawing of the self-organized faceted structure formed on the ring-shaped opening in the mask.

Figure 2. SEM images of the single (a, c) and double (b, d) GaN hexagonal microcavities. The insets in (c) and (d) show the bird-view images of the single- and double-wall hexagons, respectively.

Figure 3. Diagram of processes occurring with the growth of submicron GaN structures in shallow flat (a, b, c) and deep V- shaped (d, e, f) stripes.

Figure 4. Panchromatic CL images showing spatial CL intensity distribution in the microcavities. Bright and dark contrast corresponds to the high and to the low CL intensity, respectively. Room temperature CL images of (a) the single and (c) double hexagons. (b) and (d) - CL images detected at 10 K in the single and double hexagons, respectively.

Figure 5. CL spectra measured for (a, c) the single and (b, d) for the double hexagons, respectively at room temperature (a, b) and at 10 K (c, d). The points, where CL spectra were excited, are indicated in the inserts. Reference spectra for GaN epitaxial layers are shown by dashed lines.

Figure 6. Temperature dependence of PL in the GaN single hexagon microcavity. (a) PL spectra measured at different temperatures. Spectra are shifted vertically for clarity. Circles and squares show peak energies for the emission lines obtained from fitting using Gauss functions. (b) Energy peak positions for the emission lines in the GaN microcavity (closed signs) as a function of temperature are compared with thermal behavior of GaN exciton emission (open triangles) obtained for the GaN layer.

Figure 7. Calculations showing the distribution of the electromagnetic field intensity of the modes inside the ideal hexagonal resonators. (a) Strong localized mode experiences a total internal reflection on the edges of the single hexagon, where the maximum intensity of the standing electromagnetic wave is shifted to the corner of the hexagon. (b) Example of a weakly localized mode in the single hexagon resonator with intensity more uniformly distributed within the structure. Calculations performed for the double hexagon resonator show (c) a weakly localized mode for the outer hexagon and (d) strongly localized mode for the inner hexagon, respectively.

Figure 8. Energies and quantities $F$ for the modes localized in the single (a) and double (b) 2D hexagonal structures, respectively. The 2D resonators are shown in Figure 7. Histograms illustrating distribution of cavity modes depending on the quantity $F$ for single (c) and double (d) hexagonal structures, respectively. Dependence of the excitonic contribution to the polaritonic state $|a_{0i}|^2$ as a function of the energy of the polariton mode for the single (e) and double (f) hexagon microcavity, respectively.

Figure 9. Model geometry used in the simulations. In the left is model for single-wall hexagonal microcavity. In the right is model for double-wall hexagonal microcavity.

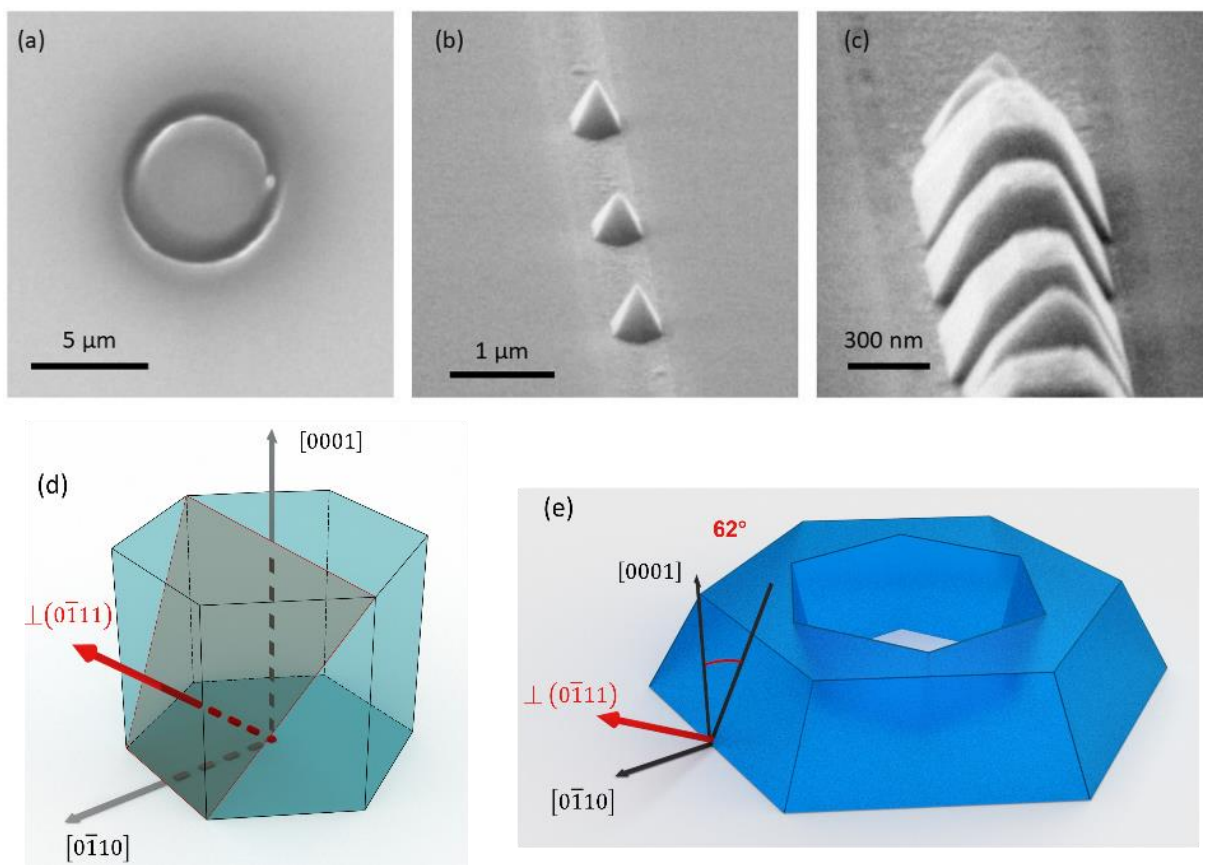

**Figure 1**

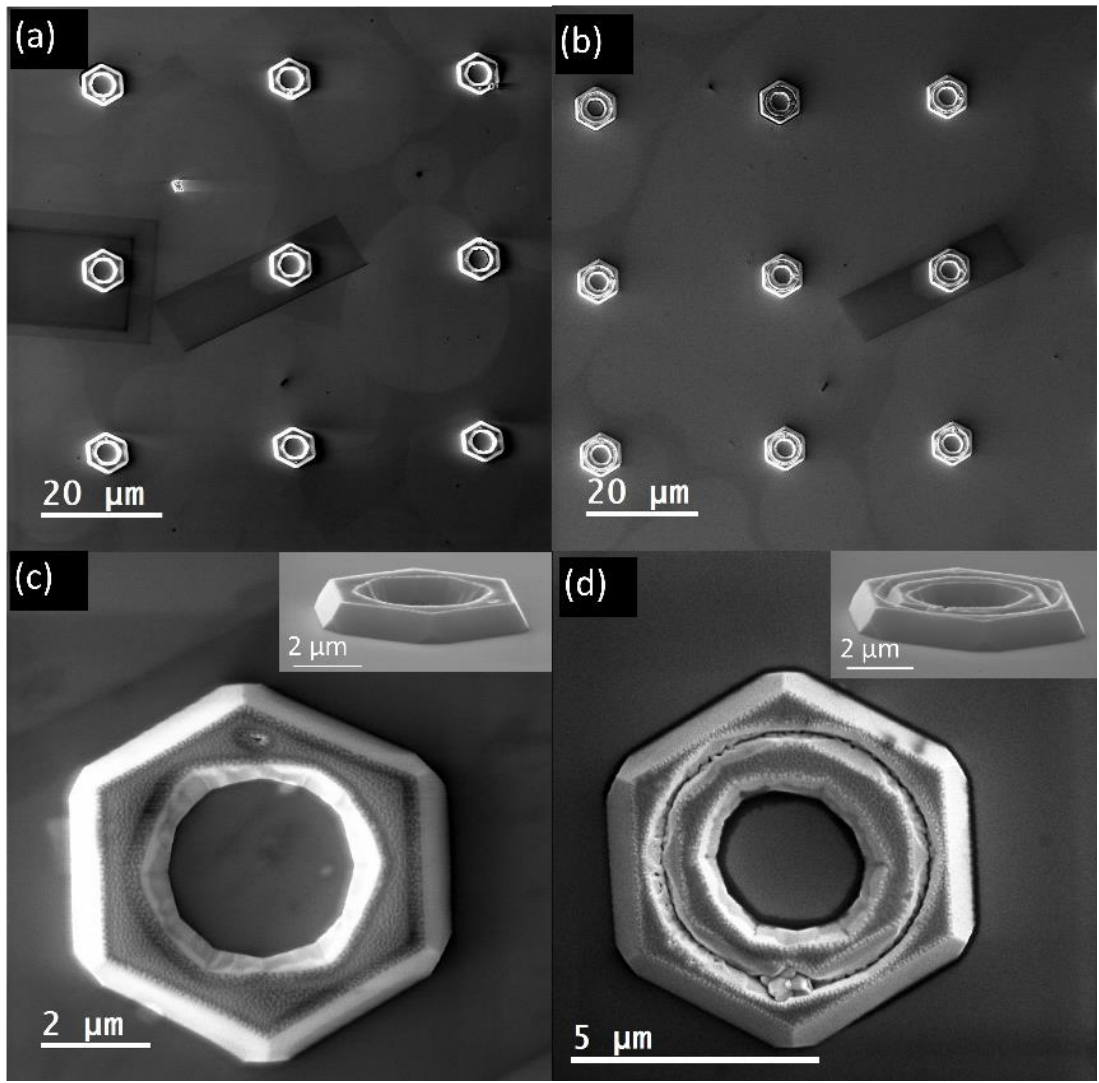

**Figure 2**

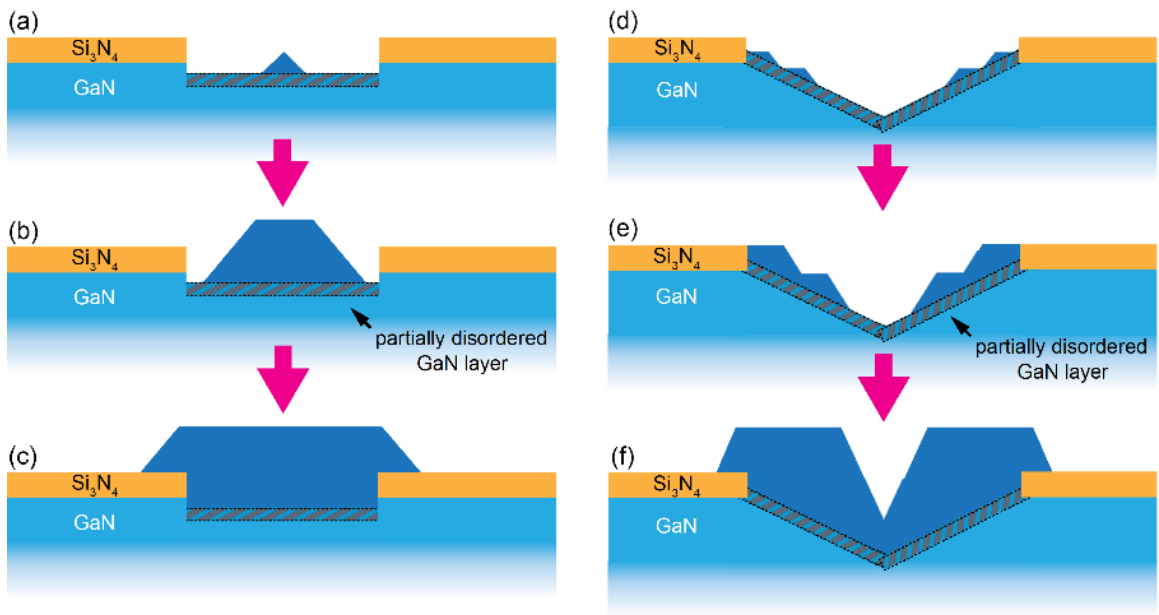

**Figure 3**

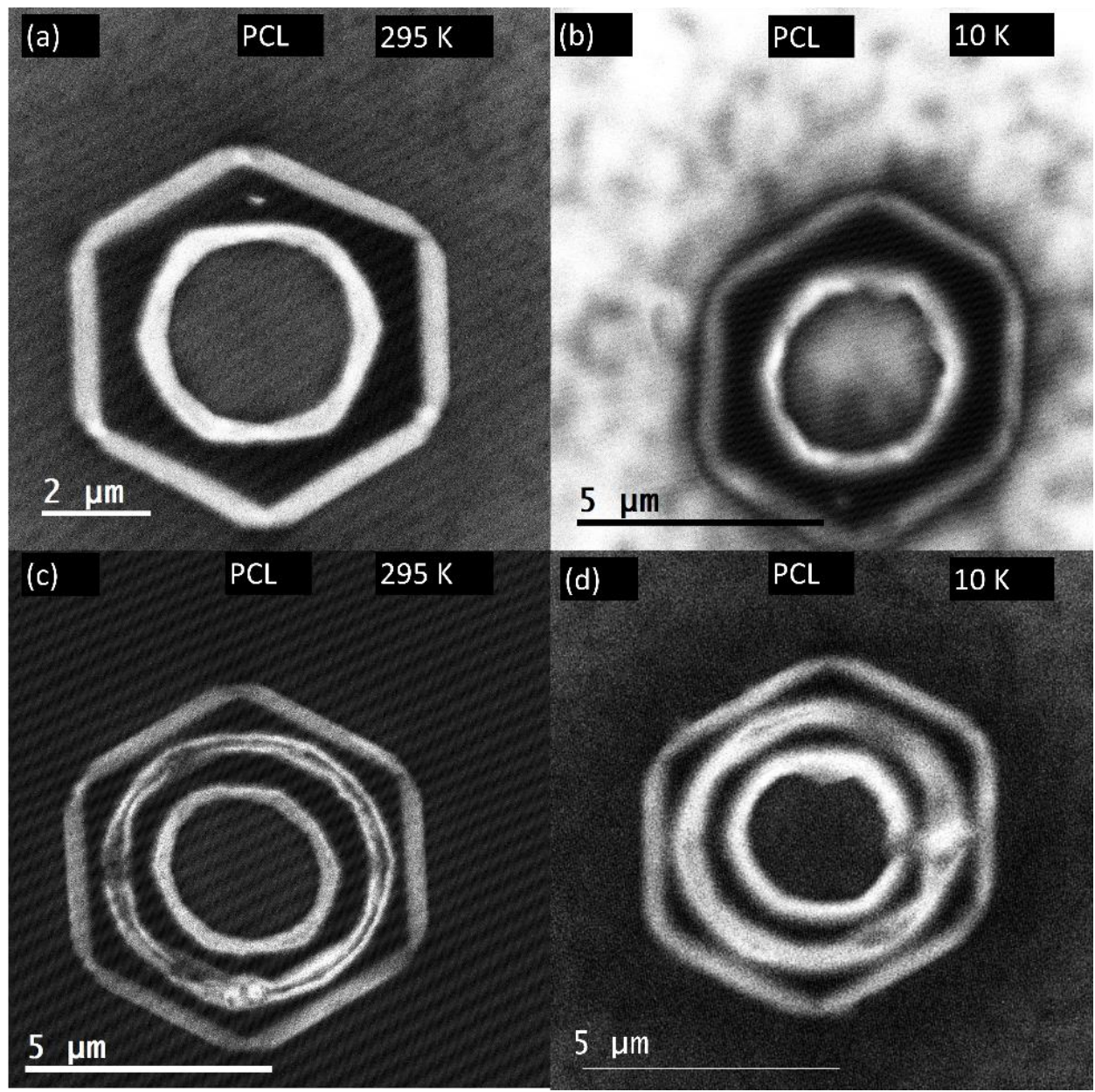

**Figure 4**

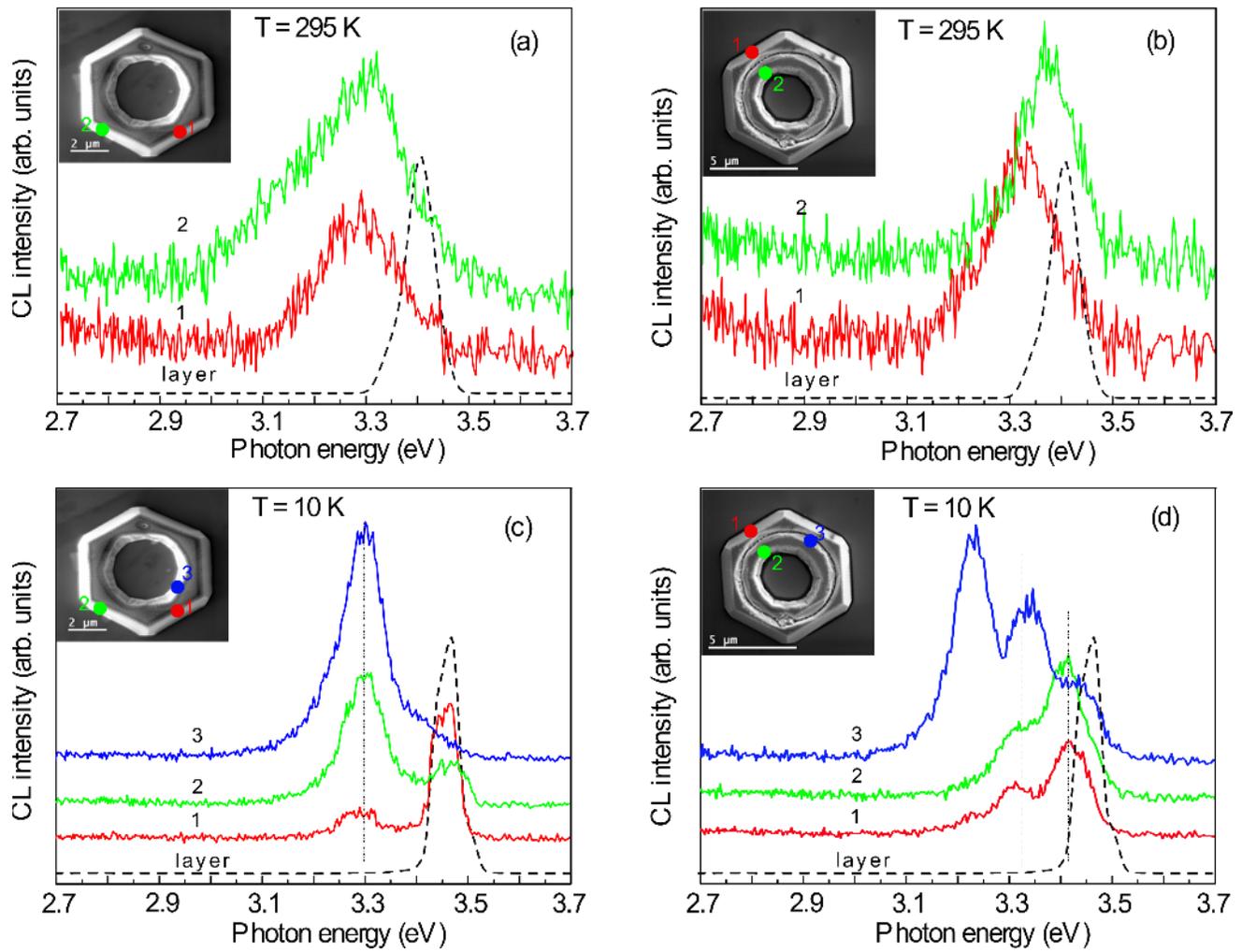

**Figure 5**

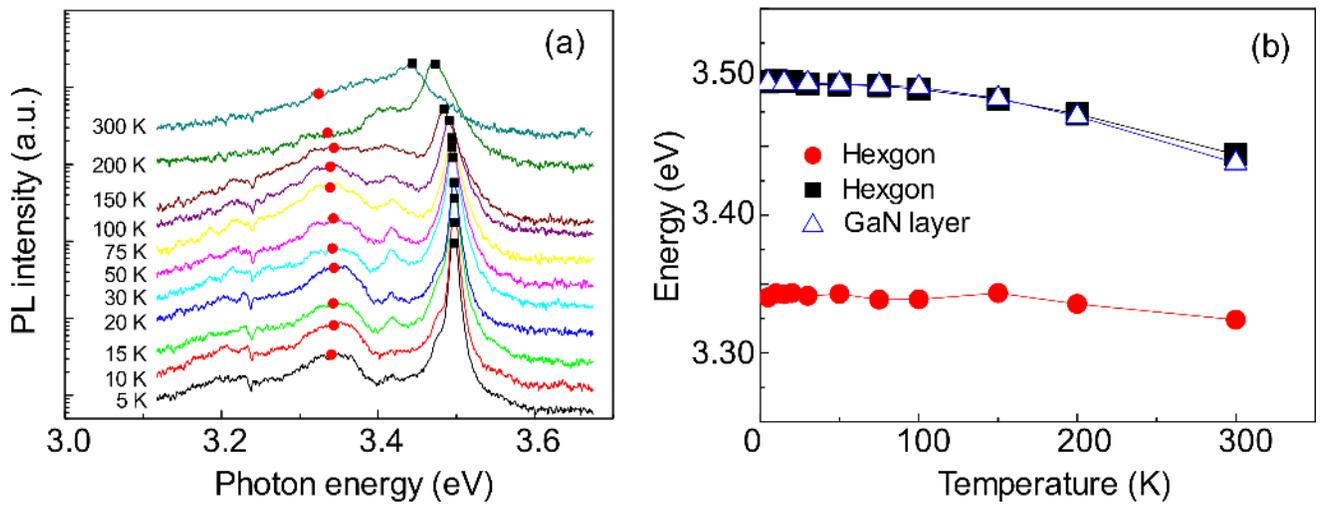

**Figure 6**

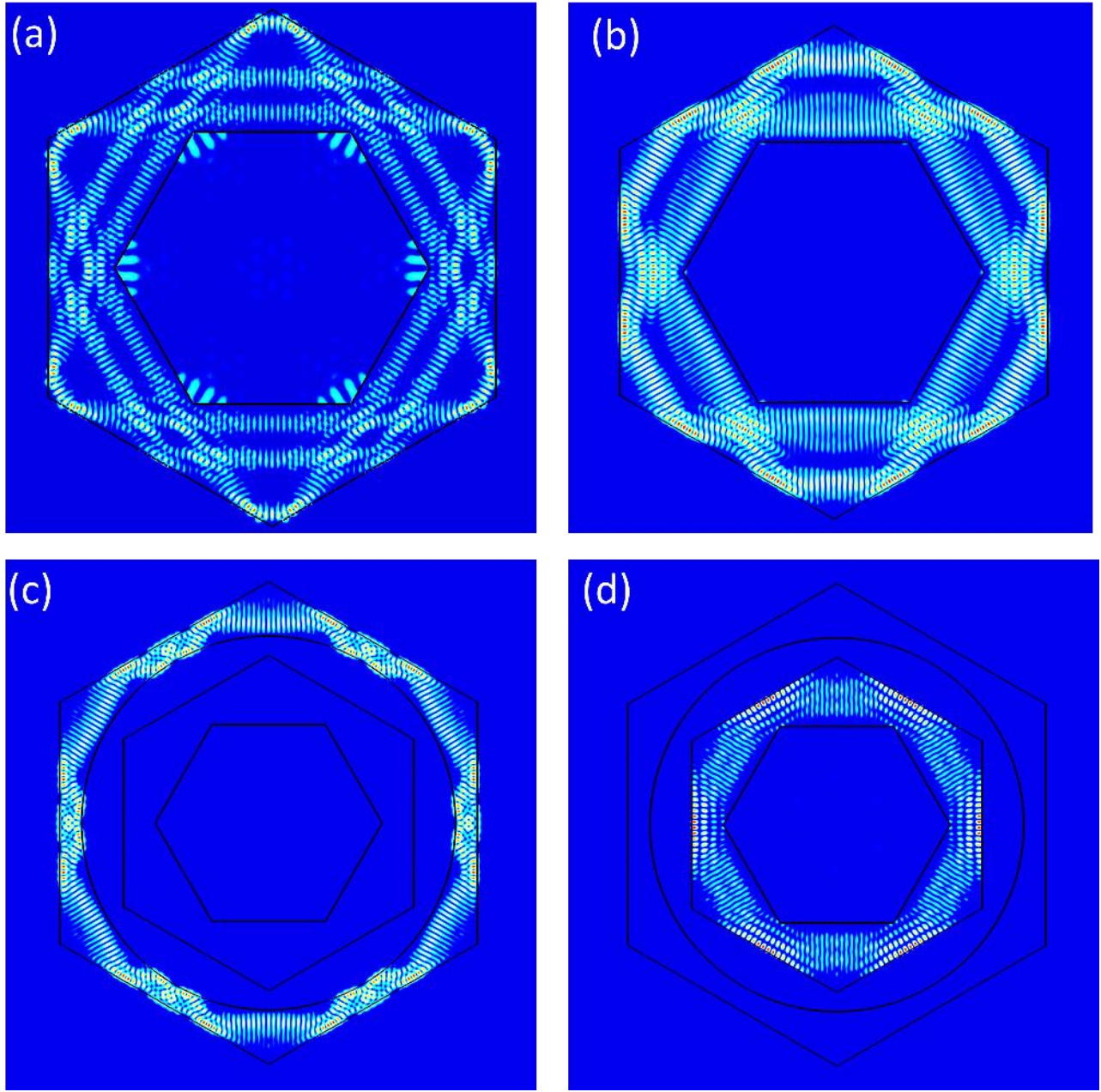

**Figure 7**

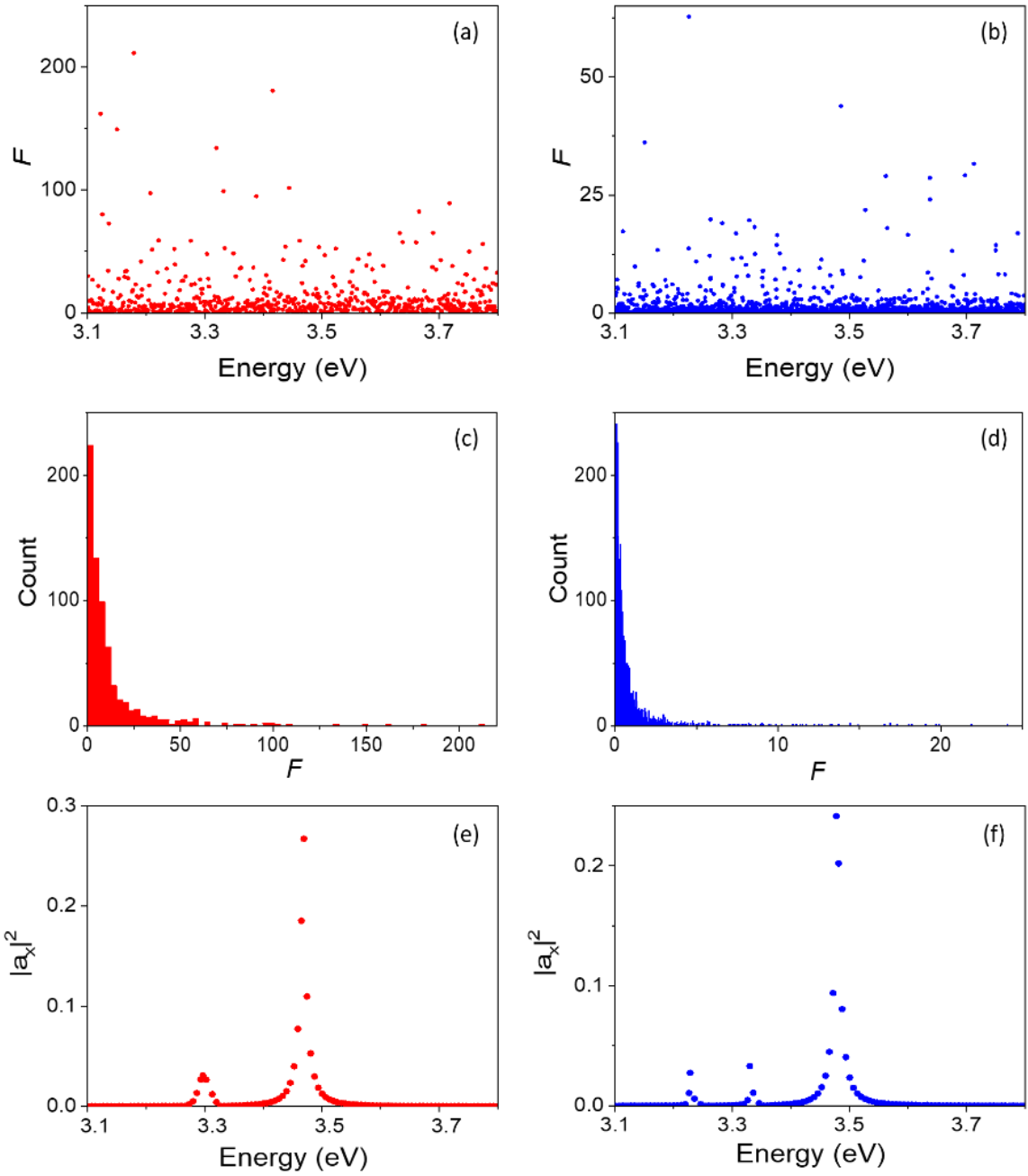

**Figure 8**

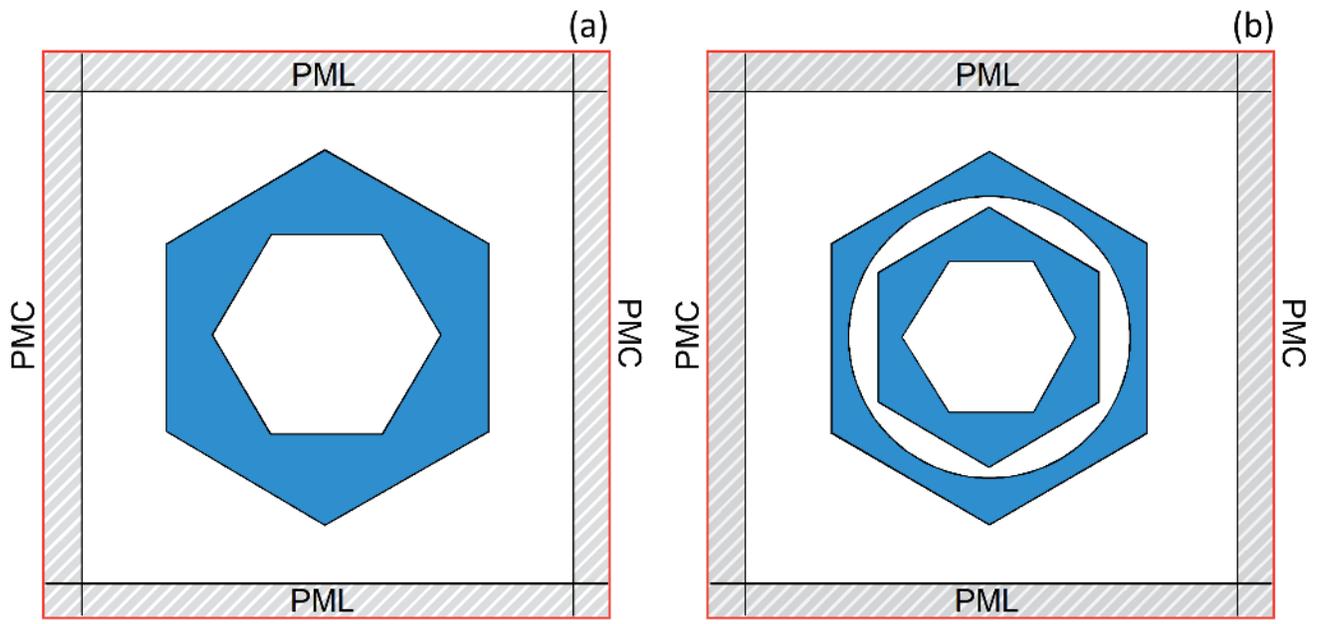

**Figure 9**